\documentclass[prb,a4paper,12pt,onecolumn,superscriptaddress,amsmath,amsfonts,amssymb,preprintnumbers,citeautoscript]{revtex4}

\usepackage[pdftex]{graphicx}
\usepackage{color}
\usepackage{ulem}

\begin{document}

\title{Long-range charge density wave proximity effect at cuprate-manganate interfaces}

\author{A.~Frano\footnote{Present Address: Materials Sciences Division, Lawrence Berkeley National Laboratory, Berkeley, California 94720, USA}}
\affiliation{Max-Planck-Institut~f\"{u}r~Festk\"{o}rperforschung, Heisenbergstr.~1, D-70569 Stuttgart, Germany}
\affiliation{Helmholtz-Zentrum Berlin f\"{u}r Materialien und Energie, Wilhelm-Conrad-R\"{o}ntgen-Campus BESSY II, Albert-Einstein-Str. 15, D-12489 Berlin, Germany}

\author{S.~Blanco-Canosa\footnote{Present Address: CIC nanoGUNE, 20018 Donostia-San Sebastian, Basque Country, Spain}}
\affiliation{Max-Planck-Institut~f\"{u}r~Festk\"{o}rperforschung, Heisenbergstr.~1, D-70569 Stuttgart, Germany}

\author{E.~Schierle}
\affiliation{Helmholtz-Zentrum Berlin f\"{u}r Materialien und Energie, Wilhelm-Conrad-R\"{o}ntgen-Campus BESSY II, Albert-Einstein-Str. 15, D-12489 Berlin, Germany}

\author{Y.~Lu}
\affiliation{Max-Planck-Institut~f\"{u}r~Festk\"{o}rperforschung, Heisenbergstr.~1, D-70569 Stuttgart, Germany}

\author{M.~Wu}
\affiliation{Max-Planck-Institut~f\"{u}r~Festk\"{o}rperforschung, Heisenbergstr.~1, D-70569 Stuttgart, Germany}

\author{M.~Bluschke}
\affiliation{Max-Planck-Institut~f\"{u}r~Festk\"{o}rperforschung, Heisenbergstr.~1, D-70569 Stuttgart, Germany}
\affiliation{Helmholtz-Zentrum Berlin f\"{u}r Materialien und Energie, Wilhelm-Conrad-R\"{o}ntgen-Campus BESSY II, Albert-Einstein-Str. 15, D-12489 Berlin, Germany}

\author{M.~Minola}
\affiliation{Max-Planck-Institut~f\"{u}r~Festk\"{o}rperforschung, Heisenbergstr.~1, D-70569 Stuttgart, Germany}

\author{G.~Christiani}
\affiliation{Max-Planck-Institut~f\"{u}r~Festk\"{o}rperforschung, Heisenbergstr.~1, D-70569 Stuttgart, Germany}

\author{H.~U.~Habermeier}
\affiliation{Max-Planck-Institut~f\"{u}r~Festk\"{o}rperforschung, Heisenbergstr.~1, D-70569 Stuttgart, Germany}

\author{G.~Logvenov}
\affiliation{Max-Planck-Institut~f\"{u}r~Festk\"{o}rperforschung, Heisenbergstr.~1, D-70569 Stuttgart, Germany}

\author{Y. Wang}
\affiliation{Max-Planck-Institut~f\"{u}r~Festk\"{o}rperforschung, Heisenbergstr.~1, D-70569 Stuttgart, Germany}

\author{P. A. van Aken}
\affiliation{Max-Planck-Institut~f\"{u}r~Festk\"{o}rperforschung, Heisenbergstr.~1, D-70569 Stuttgart, Germany}

\author{E.~Benckiser}
\affiliation{Max-Planck-Institut~f\"{u}r~Festk\"{o}rperforschung, Heisenbergstr.~1, D-70569 Stuttgart, Germany}

\author{E.~Weschke}
\affiliation{Helmholtz-Zentrum Berlin f\"{u}r Materialien und Energie, Wilhelm-Conrad-R\"{o}ntgen-Campus BESSY II, Albert-Einstein-Str. 15, D-12489 Berlin, Germany}

\author{M.~Le~Tacon}
\affiliation{Max-Planck-Institut~f\"{u}r~Festk\"{o}rperforschung, Heisenbergstr.~1, D-70569 Stuttgart, Germany}

\author{B.~Keimer}
\affiliation{Max-Planck-Institut~f\"{u}r~Festk\"{o}rperforschung, Heisenbergstr.~1, D-70569 Stuttgart, Germany}

\maketitle

\textbf{The interplay between charge density waves (CDWs) and high-temperature superconductivity is currently under intense investigation~\cite{Wu2011,Ghiringhelli2012, Chang2012, Achkar2012, Blackburn2013, Blanco2013, Silva2014, Comin2014, Hayward2014, Tabis2014}. Experimental research on this issue is difficult because CDW formation in bulk copper-oxides is strongly influenced by random disorder~\cite{LeTacon2014, Wu2014, Nie2014}, and a long-range-ordered CDW state in high magnetic fields \cite{LeBoeuf2013, Sebastian2014,Grissonnanche2014,Gerber2015} is difficult to access with spectroscopic and diffraction probes. Here we use resonant x-ray scattering in zero magnetic field to show that interfaces with the metallic ferromagnet La$_{2/3}$Ca$_{1/3}$MnO$_3$ greatly enhance CDW formation in the optimally doped high-temperature superconductor YBa$_2$Cu$_3$O$_{6+\delta}$ ($\bf \delta \sim 1$), and that this effect persists over several tens of nm. The wavevector of the incommensurate CDW serves as an internal calibration standard of the charge carrier concentration, which allows us to rule out any significant influence of oxygen non-stoichiometry, and to attribute the observed phenomenon to a genuine electronic proximity effect. Long-range proximity effects induced by heterointerfaces thus offer a powerful method to stabilize the charge density wave state in the cuprates, and more generally, to manipulate the interplay between different collective phenomena in metal oxides.}

The recent discovery and characterization of charge density waves (CDWs) in copper-oxide superconductors~\cite{Wu2011,Ghiringhelli2012, Chang2012, Achkar2012, Blackburn2013, Blanco2013, Silva2014, Comin2014, Hayward2014, Tabis2014} has opened up new perspectives on the long-standing question whether fluctuations of a competing order parameter drive or enhance high-temperature superconductivity. While the suppression of static CDW order below the superconducting (SC) critical temperature, $T_c$, demonstrates that both ground states are competing, recent evidence suggests that fluctuating CDWs play an essential role in the mechanism of high-$T_c$ superconductivity. In particular, doping-dependent resonant x-ray scattering (RXS) studies indicate that the CDW stability range in the copper-oxide phase diagram ends at the doping level with optimal $T_c$, suggesting an important influence of quantum-critical CDW fluctuations on superconductivity~\cite{Blanco2014}. Moreover, the fluctuation regimes of SC and CDW order are almost congruent, and the CDW critical fluctuations have hence been interpreted in terms of a composite CDW-SC order parameter~\cite{Hayward2014}. However, both x-ray scattering ~\cite{LeTacon2014} and nuclear magnetic resonance \cite{Wu2014} experiments have shown that pinning to residual defects is responsible for the gradual onset of static, short-range CDW order upon cooling, thus revealing a major influence of disorder on the temperature-dependent CDW correlations even in the cleanest cuprate superconductors such as  YBa$_2$Cu$_3$O$_{6+\delta}$ (YBCO). These results agree with theoretical considerations of the influence of quenched disorder on incommensurate CDWs~\cite{Nie2014}. To fully understand the interplay between CDWs and high-$T_c$ superconductivity, it will thus be essential to systematically manipulate the structure and density of defects in cuprate superconductors, and to monitor their influence on CDW formation. 

In heterostructures and superlattices, the interplay between different electronic ground states is modulated through extended heterointerfaces, which can be systematically controlled and characterized~\cite{Mannhart2010, Hwang2012}. Electronic proximity effects at interfaces have been intensely studied in ordinary metals and superconductors, where they typically extend over length scales of several nm. In metal-oxide heterostructures, there have been various reports of much longer-range proximity effects. Examples include Josephson tunneling through a 20 nm thick La$_2$CuO$_{4+\delta}$ barrier \cite{Bozovic2004} and a metal-insulator transition in a 70 nm thick VO$_2$ film induced by charge transfer at the surface~\cite{Nakano2012}. Subsequent work has, however, pointed out the possibly crucial role of oxygen off-stoichiometry, which can modify the phase behavior of oxide heterostructures in a bulk-like manner unrelated to interface physics~\cite{Jeong2013}. It has thus far proven difficult to determine the oxygen concentration in thin-film heterostructures precisely enough to conclusively discriminate between these scenarios.

We have overcome these difficulties and were able to demonstrate a genuine, long-range interfacial proximity effect by using RXS to probe CDW correlations in well-characterized superlattices (SLs) of fully oxygenated YBCO and the metallic ferromagnet La$_{2/3}$Ca$_{1/3}$MnO$_3$ (LCMO) grown on SrTiO$_3$ substrates (see Methods and Supplementary Materials). The LCMO layer thickness was kept fixed at 10 nm, and the YBCO layer thickness was set to $D=10$, 20, and 50 nm. Both the substrate-induced strain and the mutual strain of the SL constituents are negligible and do not influence the physical properties \cite{Driza2012}. Closely related multilayer structures have served as model systems for the interplay between ferromagnetism and superconductivity, \cite{Sefrioui2003,Pena2004,Hoppler2009,Kalcheim2011,Visani2012} interfacial orbital and magnetic reconstructions, ~\cite{Hoffmann2005,Stahn2005,Chakhalian2006,Chakhalian2007,Yunoki2007,Chien2014,Satapathy2012} and electron-phonon interactions \cite{Driza2012}, but the influence of YBCO-LCMO interfaces on CDW formation has not yet been studied.

\begin{figure*}
\includegraphics[width=1\columnwidth]{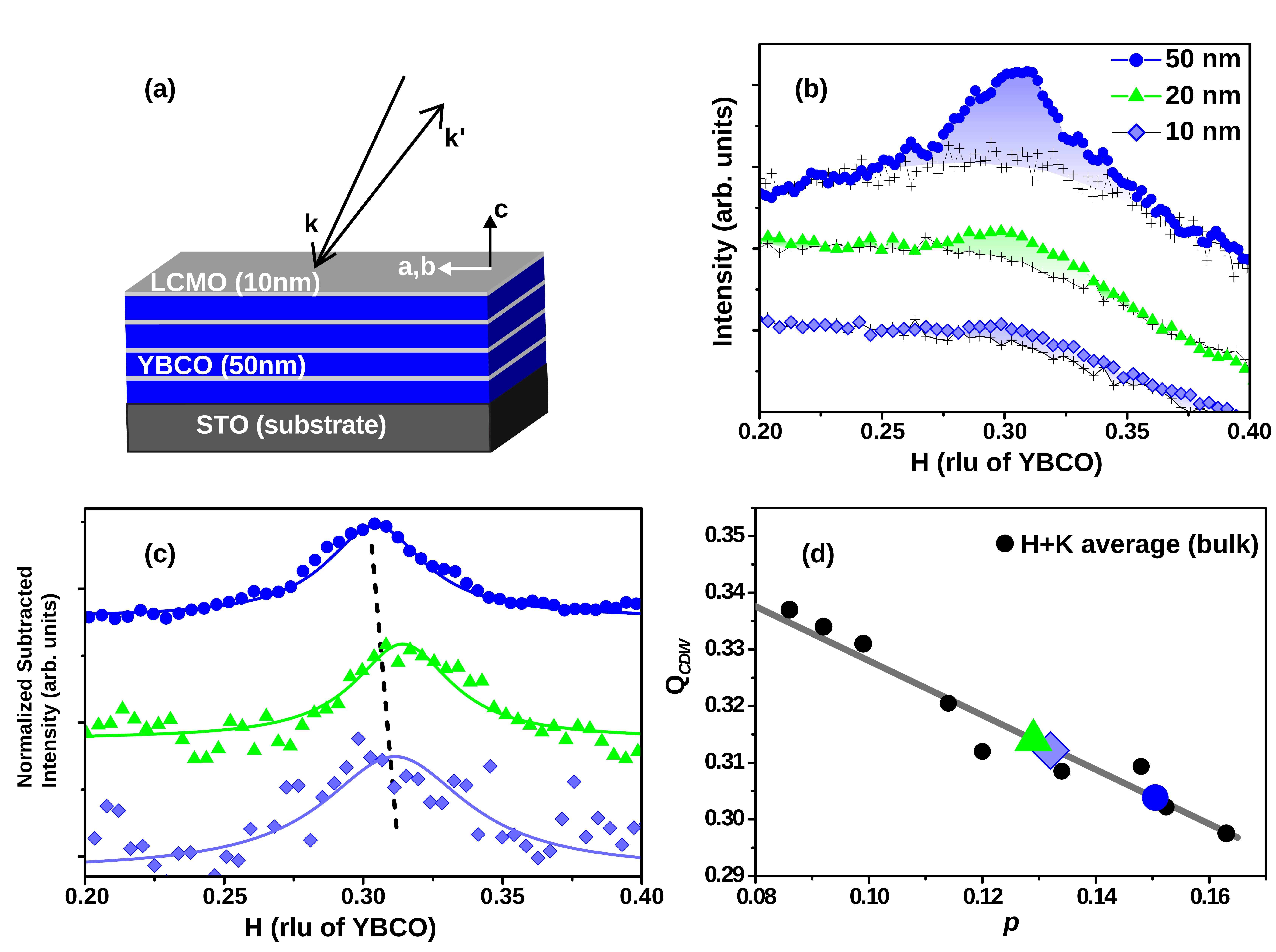}
\caption{{\bf Resonant x-ray scattering from YCBO-LCMO superlattices.} a, Sketch of the experimental geometry for one of the SLs. Layer thicknesses, incoming and outgoing x-ray wavevectors ($k$ and $k^\prime$), and crystallographic axes of YBCO ($a$, $b$, $c$) are marked. b, RXS scans along the \textbf{Q} = (1,0,0)/(0,1,0) momentum-space direction for SLs with $10,20,50$\,nm YBCO layers taken at temperature $T = 10$ K (solid symbols). The crosses show background scans at $T=180$ K, where the RXS signal is $T$-independent. All scans were taken with the same counting time. c, Background-subtracted data at $T = 10$ K. The lines are the results of fits to Lorentzian profiles. For clarity, the data were normalized to the fitted peak amplitude for each scan. d, CDW wavevector observed in the SLs compared to the \textbf{Q}$_{CDW}$-versus-$p$ relation in single crystal YBCO.~\cite{Blanco2014} Since the YBCO layers in the SLs are twinned, the average of \textbf{Q}$_{CDW}$ along the (1,0,0) and (0,1,0) direction in bulk YBCO is used for comparison. The error bars are smaller than the symbol size.}
\label{fig:figure1}
\end{figure*}

Following prior RXS work on bulk YBCO, \cite{Ghiringhelli2012,Achkar2012,Blanco2013,Blanco2014} we tuned the photon energy to the $L$-absorption edge of planar Cu, so that the resulting data become highly sensitive to the valence electron system in the CuO$_2$ planes (Fig.~\ref{fig:figure1}a). Figure~\ref{fig:figure1}b shows RXS scans for three SLs with different YBCO layer thicknesses at different temperatures.  In all three samples, well-defined incommensurate peaks develop with decreasing temperature around the planar wave vectors $\textbf{Q}_{CDW} \sim (0.3,0)$ and $(0, 0.3)$ (which cannot be distinguished in the SLs because of twinning of the orthorhombic structure). This behavior is closely analogous to the one of moderately doped bulk cuprates, where it was shown to arise from CDW correlations that grow upon cooling. \cite{Ghiringhelli2012, Chang2012, Achkar2012, Blackburn2013, Blanco2013, Silva2014, Comin2014, Hayward2014, Tabis2014} Since the YBCO layers in the SLs are fully oxygenated, however, the intense CDW peaks observed in the samples with the 20 and 50 nm thick YBCO layers are in striking contrast to the behavior of bulk YBCO, where the CDW reflections are very weak or absent at optimal doping. \cite{Blanco2014}

In bulk YBCO, the CDW wave vector $\textbf{Q}_{CDW}$ was found to decrease approximately linearly with the doping level, $p$, as expected for an instability of a hole-like Fermi surface~\cite{Blanco2014}. Taking advantage of this $\textbf{Q}_{CDW}-p$ calibration, we can use the positions of the CDW peaks (which were extracted by fitting the RXS data to Lorentzian profiles, Fig. 1c) to accurately determine the average doping level, $\langle p \rangle$, in the YBCO layers of the SLs with $D=20$ and 50 nm thick YBCO layers. (In the sample with $D=10$ nm, the error bars are larger due to the smaller signal intensity.) The plot in Fig. 1d shows that the average YBCO doping level in SLs the $D=20$ nm sample is $\langle p \rangle \sim 0.12$, placing it in the moderately underdoped regime of the phase diagram. In the $D=50$ nm sample, on the other hand, the YBCO layers are optimally doped on average ($\langle p \rangle \sim 0.15$). Note that the momentum width of the RXS reflections in the SLs is inhomogeneously broadened by averaging both over the charge density profile and over structural twin domains in the YBCO layers, so that the CDW correlation length cannot be extracted from the RXS profiles.

These data demonstrate that the average doping level of the YBCO layers increases with increasing layer thickness, consistent with prior experimental and theoretical work that demonstrated a transfer of 0.2-0.3 holes per planar Cu across the YBCO-LCMO interface~\cite{Chakhalian2007,Yunoki2007,Chien2014}. While most of this transfer takes place in the first two unit cells~\cite{Chakhalian2007,Yunoki2007,Chien2014}, RXS is sensitive to
the ``tail'' of the interfacial charge profile, which extends more than 10 nm inside the YBCO layer. As a function of distance from the interface, the YBCO layer thus replicates the entire phase diagram comprising magnetically ordered phases (evidenced by Cu magnetic moments at the interface \cite{Stahn2005,Chakhalian2006}), CDW order, and superconductivity (Fig. 2).

Beyond this qualitative analogy, the intensity of the RXS reflections reveals striking differences between the CDW correlations in the SLs and in the bulk. Whereas in bulk optimally doped YBCO the RXS reflections characteristic of CDW order are extremely weak~\cite{Blanco2014}, their intensity in the SLs grows with increasing YBCO layer thickness (and hence increasing $\langle p \rangle$). Note that the LCMO layers are nearly transparent to x-rays in resonance to the Cu $L$-absorption edge, and that the YBCO volume probed in all three samples was comparable (see Methods). The systematic growth of the CDW peak intensity with thickness and its large intensity for the $D=50$ nm sample demonstrate that most (if not all) of the YBCO volume in this SL is affected by CDW formation. Rather than being pinned to the interfaces, as expected for an ordinary proximity effect, these data imply that robust CDW order is present over a large fraction of the 50 nm thick layer with $\langle p \rangle = 0.15$.

Having established the presence of robust CDW order in the 50 nm thick YBCO layer with $\langle p \rangle = 0.15$, we now turn to its temperature and magnetic field dependence. The temperature dependence of the RXS intensity (Fig. 3) is indicative of a second-order phase transition with a critical temperature of 110 K. This is in stark contrast to the gradual onset of CDW correlations with decreasing temperature in bulk cuprates (shown for comparison in Fig. 3), which has been attributed to the competition between CDW and superconductivity and/or pinning of CDW domains to random defects~\cite{LeTacon2014,Hayward2014}. The RXS intensity in the SLs evolves smoothly through the superconducting transition, with no sign of the sharp suppression below $T_c$ seen in bulk YBCO. Moreover, Figure 4 shows that a magnetic field of 6 T does not affect the CDW correlations, again in contrast to the behavior of bulk underdoped YBCO where the CDW is markedly enhanced by magnetic fields of this magnitude \cite{Chang2012,Blanco2013,Blanco2014}.

These observations indicate that the CDW state in YBCO-LCMO superlattices is much closer to a genuine thermodynamic phase than it is in bulk YBCO. This provides a natural explanation for modifications of the electron-phonon interactions \cite{Driza2012} and the thermoelectric properties \cite{Heinze2012} over a similar spatial range. Different mechanisms may contribute to the stabilization of the CDW over a range of tens of nm. In particular, we note that the graded charge carrier concentration profile (Fig. 2) includes regions close to the interface where $p$ is optimal for the formation of the CDW \cite{Blanco2014}. These regions can act effectively as coherent nucleation centers of CDW domains in optimally doped regions further inside the YBCO layers. In contrast, pinning of incommensurate CDW fluctuations by randomly disordered defects in bulk YBCO~\cite{LeTacon2014,Wu2014} is presumably much less effective.

We now discuss the proximity-induced monotonic evolution of the CDW order parameter below $T_c$. The strong, systematic increase of both the CDW peak intensity and the superconducting $T_c$ with YBCO layer thickness implies that CDW order and superconductivity coexist deep inside the YBCO layers. We therefore consider possible scenarios for laterally modulated structures comprising both superconducting and CDW order at optimum doping. The first scenario involves mesoscopic patches of non-superconducting CDW order that coexist laterally with patches of superconducting order. The order in the CDW patches is then closely related to the CDW state realized in bulk YBCO$_7$ in magnetic fields of order 100 T, where superconductivity is obliterated by orbital depairing~\cite{LeBoeuf2013,Sebastian2014,Grissonnanche2014,Gerber2015}. The superconducting patches, on the other hand, are CDW-free, as they are in bulk optimally doped YBCO.  Due to the mesoscopic phase separation, the interaction between the two order parameters is strongly reduced, thus explaining the lack of suppression of the CDW order parameter below $T_c$ (Fig. 3). However, there is no direct evidence for such mesoscopic phase separation, and the mechanisms that might give rise to such behavior remain unclear.

Second, superconductivity and CDW order may be microscopically ``intertwined''. Since CDW order is strengthened by proximity to the interfaces and is fully established at the superconducting $T_c$, the superconducting order parameter has to adjust to the pre-existing CDW order, perhaps by forming a modulated state akin to the ``Fulde-Ferrell-Larkin-Ovchinnikov'' state in ferromagnetic superconductors. Charge transfer across the interface and proximity to the ferromagnetic LCMO layers may stabilize additional order parameters (such as incommensurate antiferromagnetism or triplet superconductivity \cite{Kalcheim2011,Visani2012}) that are admixed into the composite order in the YBCO layers. This may further weaken singlet d-wave superconductivity relative to the CDW.

Both scenarios are equally intriguing. In the first case, interfaces stabilize a CDW state that is realized in bulk cuprates only in magnetic fields of order 100 T. Experiments on interface-enhanced CDWs do not require extreme conditions and hence open up novel perspectives for a complete characterization of the electronic structure associated with CDW formation, including photoemission experiments capable of imaging the full Fermi surface. In the second scenario, a new form of ``intertwined'' order is realized in an optimally doped high-temperature superconductor. Clearly, discriminating between these scenarios will require further experiments. It will also be interesting to investigate the possible contribution of CDW correlations to the anomalous transport properties of cuprate heterostructures \cite{Bozovic2004,Kalcheim2011,Visani2012}. Finally, the ``tomographic'' reconstruction of charge density profiles we have demonstrated for YBCO-LCMO may also provide novel insights into the interfacial variation of incommensurate spin, charge, and orbital correlations in other metal-oxide heterostructures \cite{Mannhart2010,Hwang2012}.

\bigskip

\noindent {\bf Methods.} The RXS experiments were performed in the UHV diffractometer at the UE46-PGM1 beamline of the Helmholtz Zentrum Berlin \cite{Fink2013}, working with vertically polarized photons in a horizontal scattering geometry. To access the $\textbf{Q}_{CDW}$ reflection in (001)-oriented samples, the asymmetric scattering geometry of Fig.~\ref{fig:figure1}a was used.
Magnetic field dependent measurements were performed in the high-field diffractometer at the same beamline.  The field was oriented at an angle of 11$^{\circ}$ from the $c$-axis. We investigated superlattices comprising $D$\,nm YBCO\,-\,10\,nm LCMO ($D=10, 20, 50$), which were grown using pulsed laser deposition on single crystalline SrTiO$_3$ substrates. A working temperature of 730$^{\circ}$C and an oxygen partial pressure of 0.5\,mbar were used to preserve the desired stoichiometry.  Structural characterization of the same samples studied here with non-resonant Cu K$_{\alpha}$ x-rays~\cite{Driza2012} yielded the room-temperature $c$-axis lattice parameter 11.695(5)\,{\AA}, close to that of YBCO with oxygen content $\delta=1$~\cite{Liang2006}. The oxygen stoichiometry was also verified by Raman measurements of the A$_{1g}$ apical oxygen mode of YBCO, as described in Ref. \, \onlinecite{Driza2012}. The superconducting transition temperatures ($T_c=45,60,82$\,K, for $D=10, 20, 50$) were determined by magnetometry~\cite{Driza2012}. The overall thickness of the multilayer stacks (15 bilayer repetitions for $D=10$, 10 for $D=20$, and 5 for $D=50$ nm) was larger than the x-ray penetration depth ($\sim 0.1 \mu$m at resonance). The interface between the substrate and the multilayer therefore does not contribute to the RXS intensity, and the YBCO volume illuminated by the x-ray beam was comparable for all samples.

\bigskip

\noindent {\bf Acknowledgments.} We acknowledge fruitful discussions with V. Hinkov, V. Zabolotnyy, A. Charnukha, G. Sawatzky, and C. Bernhard. This work was partly funded by the Deutsche Forschungsgemeinschaft within the framework of the SFB/TRR 80.

\clearpage

\begin{figure}[h]
\includegraphics[width=0.8\columnwidth]{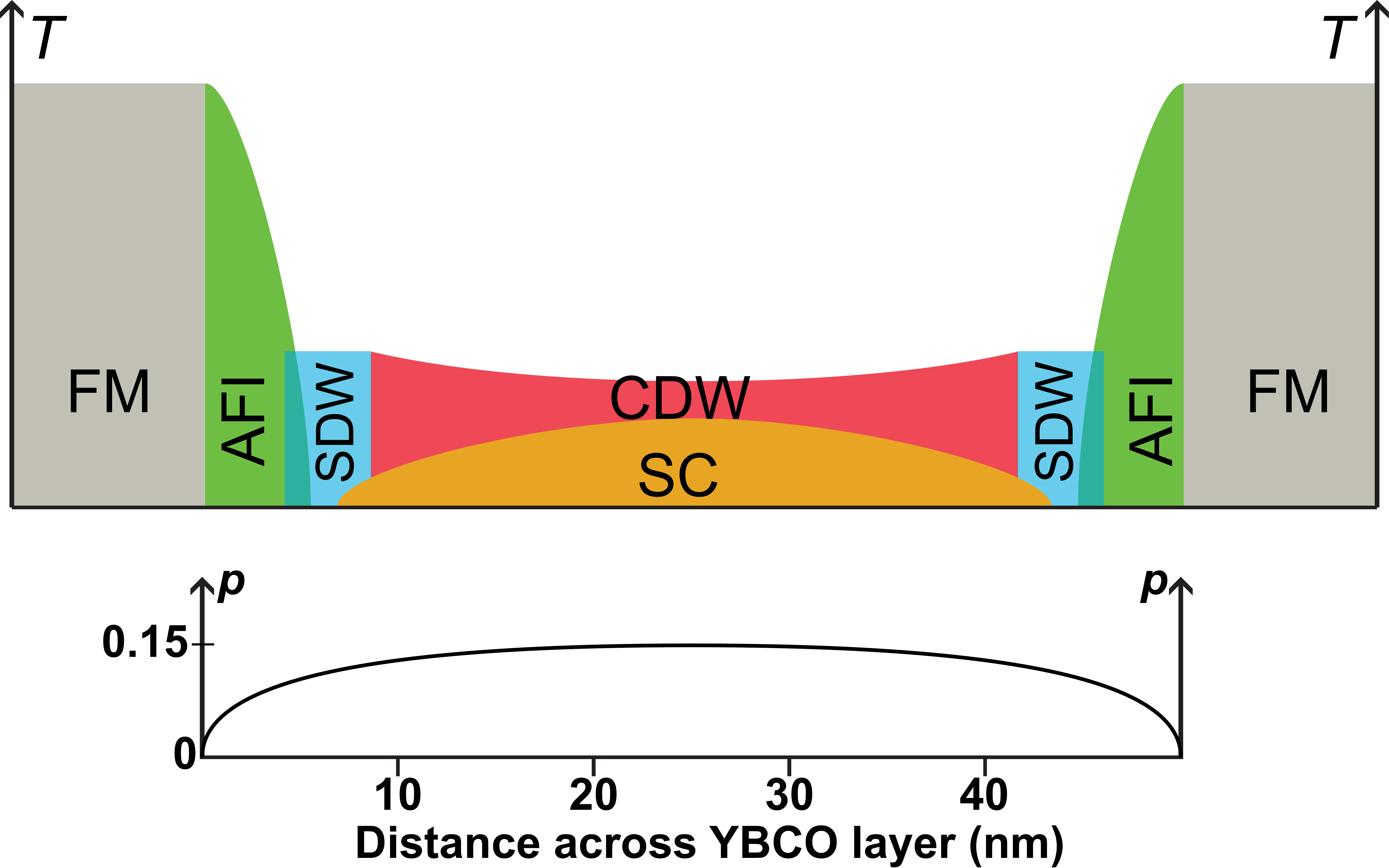}
\caption{{\bf Tomographic view of a 50 nm thick YBCO layer in a YBCO-LCMO SL.} a, Electronic phases as a function of depth and temperature including antiferromagnetic insulating (AFI), spin density wave (SDW), superconducting (SC), and charge density wave (CDW) states.  The bottom panel is an estimate of the corresponding charge carrier concentration $p$. Detailed models of the charge carrier profile will have to consider the work function difference between YBCO and LCMO, the interfacial structure, and chemical intermixing (see Supplementary Materials).}
\label{fig:figure2}
\end{figure}

\begin{figure}[h]
\includegraphics[width=0.8\columnwidth]{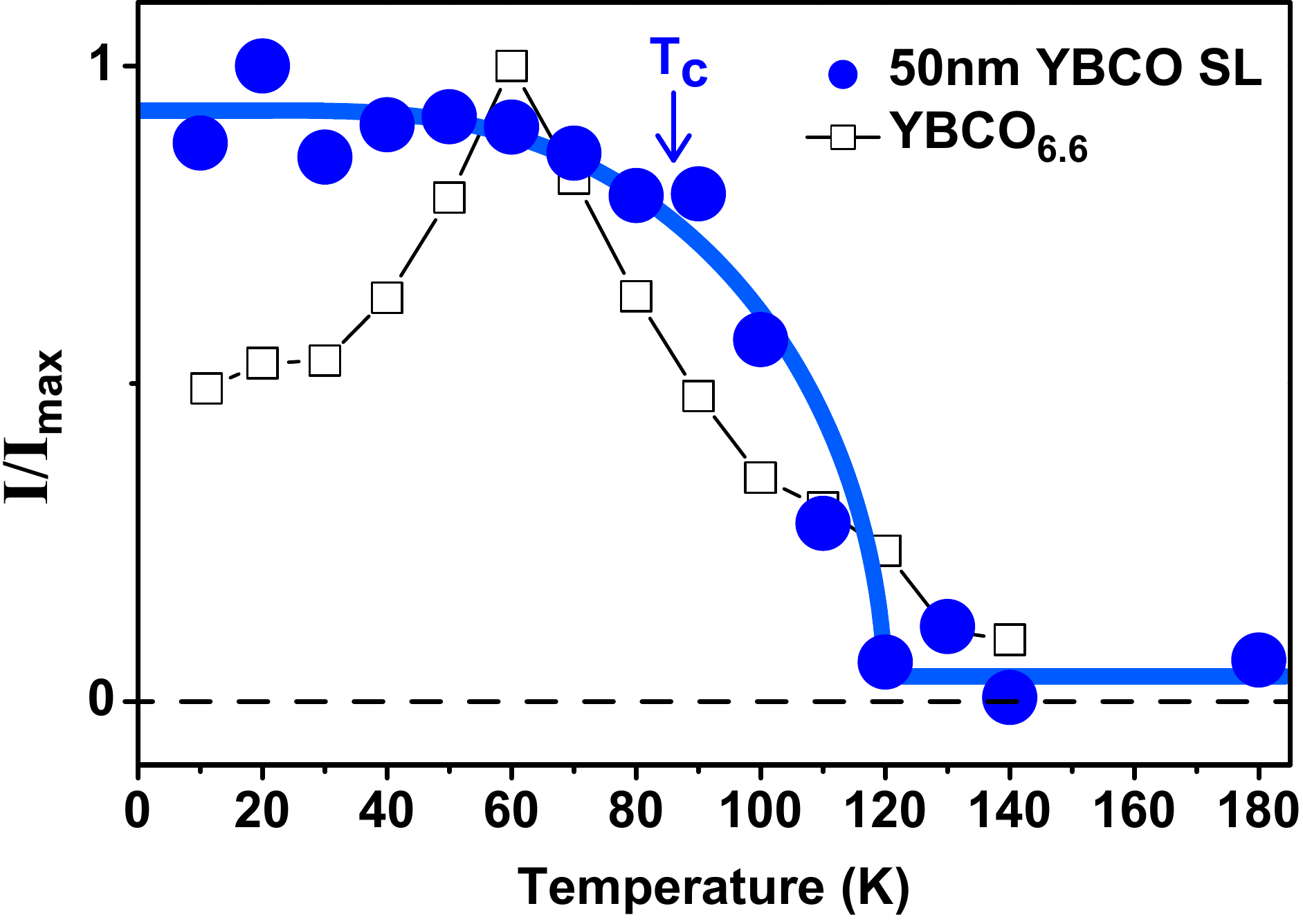}
\caption{{\bf Temperature dependence of the RXS intensity for a SL with 50 nm thick YBCO.} The data for the SL are shown as blue circles and compared to equivalent data on a single crystal of YBCO$_{6.6}$ \cite{Blanco2014}. Lines are guides-to-the-eye.}
\label{fig:Temperature_dep}
\end{figure}

\begin{figure}[h]
\includegraphics[width=0.8\columnwidth]{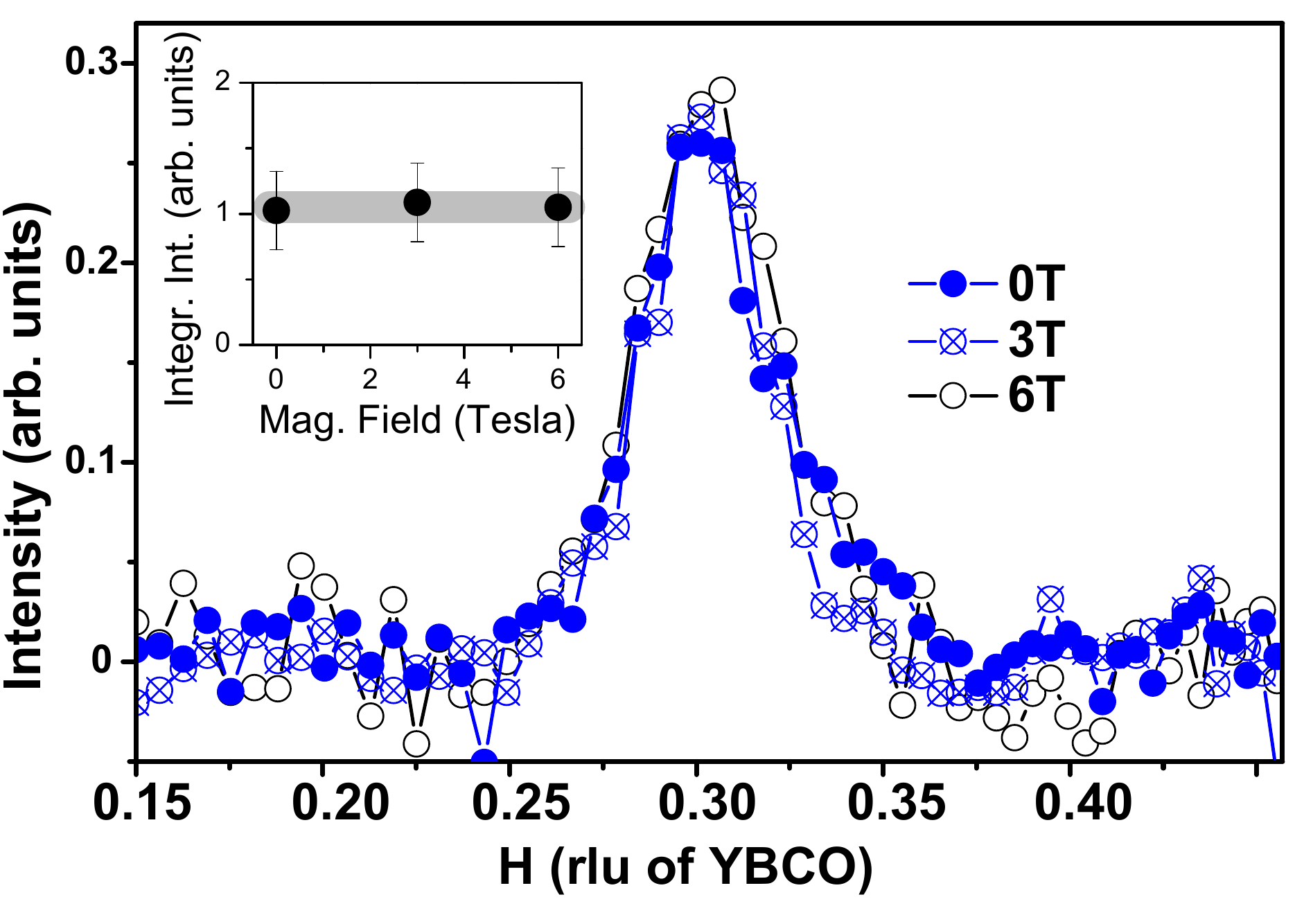}
\caption{{\bf Magnetic field dependence of the RXS intensity for a SL with 50 nm thick YBCO.} The main panel shows background-subtracted RXS scans with applied magnetic field nearly along the $c$-axis, taken at $T = 4$ K. The inset shows the magnetic field dependence of the RXS intensity, extracted from the RXS profiles by fitting to Lorentzians.}
\label{fig:figure4}
\end{figure}


\bigskip



\begin{thebibliography}{99}


\bibitem{Wu2011} T. Wu, H. Mayaffre, S. Kr\"{a}mer, M. Horvatic, C. Berthier, W. N. Hardy, R. Liang, D. A. Bonn, and M.-H. Julien, \textit{Nature} \textbf{477}, 191-194 (2011).

\bibitem{Ghiringhelli2012} Ghiringhelli, G. \textit{et al.} Long-Range Incommensurate Charge Fluctuations in (Y,Nd)Ba$_2$Cu$_3$O$_{6+x}$. \textit{Science} \textbf{337}, 821-825 (2012).

\bibitem{Chang2012} Chang, J. \textit{et al.} Direct observation of competition between superconductivity and charge density wave order in YBa$_2$Cu$_3$O$_{6.67}$.  \textit{Nature Phys.} \textbf{8}, 871-876 (2012).

\bibitem{Achkar2012} Achkar, A. J. \textit{et al.}  Distinct Charge Orders in the Planes and Chains of Ortho-III-Ordered YBa$_2$Cu$_3$O$_{6+\delta}$
Superconductors Identified by Resonant Elastic X-ray Scattering. \textit{Phys. Rev. Lett.} \textbf{109}, 167001 (2012).

\bibitem{Blackburn2013} Blackburn, E. \textit{et al.} X-Ray Diffraction Observations of a Charge-Density-Wave Order in Superconducting Ortho-II
￼YBa$_2$Cu$_3$O$_{6.54}$ Single Crystals in Zero Magnetic Field. \textit{Phys. Rev. Lett.} \textbf{110}, 137004 (2013).

\bibitem{Blanco2013} Blanco-Canosa, S. \textit{et al.} Momentum-Dependent Charge Correlations in YBa$_2$Cu$_3$O$_{6+\delta}$ Superconductors Probed
by Resonant X-Ray Scattering: Evidence for Three Competing Phases. \textit{Phys. Rev. Lett.} \textbf{110}, 187001 (2013).

\bibitem{Silva2014} da Silva Neto, E. H. \textit{et al.} Ubiquitous Interplay Between Charge Ordering and High-Temperature Superconductivity in Cuprates. \textit{Science} \textbf{343}, 393-396 (2014).

\bibitem{Comin2014}  Comin, R. \textit{et al.} Charge Order Driven by Fermi-Arc Instability in Bi$_2$Sr$_{2-x}$La$_x$CuO$_{6+\delta}$. \textit{Science} \textbf{343}, 390-392 (2014).

\bibitem{Tabis2014} Tabis, W. \textit{et al.} Charge order abd its connection with Fermi-liquid charge transport in a pristine high-$T_c$ cuprate. \textit{Nature Comm.} \textbf{5}, 5875 (2014).

\bibitem{Hayward2014} Hayward, L. E. \textit{et al.} Angular Fluctuations of a Multicomponent Order Describe the Pseudogap of YBa$_2$Cu$_3$O$_{6+x}$. \textit{Science} \textbf{343}, 1336-1339 (2014).

\bibitem{LeTacon2014} Le Tacon, M. \textit{et al.} Inelastic X-ray scattering in YBa$_2$Cu$_3$O$_{6.6}$ reveals giant phonon anomalies and elastic central peak due to charge-density-wave formation. \textit{Nature Phys.} \textbf{10}, 52-58 (2014).

\bibitem{Wu2014} Wu, T. \textit{et al.} Short-range charge order reveals the role of disorder in the pseudogap state of high-T$_c$ superconductors. arXiv:1404.1617.


\bibitem{Nie2014} Nie, L., Tarjus, G. \& Kivelson, S. A. Quenched disorder and vestigial nematicity in the pseudogap regime of the cuprates.
\textit{Proc. Nat. Acad. Sci. (USA)} \textbf{111}, 7980-7985 (2014).

\bibitem{LeBoeuf2013} Le Boeuf, D. \textit{et al.} Thermodynamic phase diagram of static charge order in underdoped YBa$_2$Cu$_3$O$_{y}$.  \textit{Nature Phys.}  \textbf{9}, 79-83 (2013).

\bibitem{Sebastian2014} Sebastian, S. E. \textit{et al.} Normal-state electronic structure in underdoped high-$T_c$ copper oxides.  \textit{Nature} \textbf{511}, 61-64 (2014).

\bibitem{Grissonnanche2014} Grissonnanche, G. \textit{et al.} Direct measurement of the upper critical field in cuprate superconductors. \textit{Nature Commun.} \textbf{5}, 3280 (2014).

\bibitem{Gerber2015} Gerber, S. \textit{et al.} Three-dimensional charge density wave order in YBa$_2$Cu$_3$O$_{6.67}$ at high magnetic fields. \textit{Science} \textbf{350}, 949-952 (2015).

\bibitem{Blanco2014} Blanco-Canosa, S. \textit{et al.} Resonant x-ray scattering study of charge-density wave correlations in YBa$_2$Cu$_3$O$_{6+x}$. \textit{Phys. Rev. B} \textbf{90}, 054513 (2014).

\bibitem{Mannhart2010} Mannhart, J. \& Schlom, D. G. Oxide interfaces: an opportunity for electronics. \textit{Science} \textbf{327}, 1607-1611 (2010).

\bibitem{Hwang2012} Hwang, H. Y. \textit{et al.} Emergent phenomena at oxide interfaces. \textit{Nature Mater.} \textbf{11}, 103-113 (2012).

\bibitem{Bozovic2004} Bozovic, I. \textit{et al.}
Giant Proximity Effect in Cuprate Superconductors. \textit{Phys. Rev. Lett.} \textbf{93}, 157002 (2004).

\bibitem{Nakano2012} Nakano, M. \textit{et al.} Collective bulk carrier delocalization driven by electrostatic surface charge accumulation. \textit{Nature} \textbf{487}, 459-462 (2012).

\bibitem{Jeong2013} Jeong, J. \textit{et al.} Suppression of Metal-Insulator Transition in VO$_2$ by Electric Field Induced Oxygen Vacancy Formation. \textit{Science} \textbf{339}, 1402-1405 (2013).

\bibitem{Driza2012} Driza, N. \textit{et al.} Long-range transfer of electron-phonon coupling in oxide superlattices. \textit{Nature Mater.}  \textbf{11}, 675-681 (2012).

\bibitem{Sefrioui2003} Sefrioui, Z. \textit{et al.} Ferromagnetic/superconducting proximity effect in La$_{0.7}$Ca$_{0.3}$MnO$_{3}$/YBa$_{2}$Cu$_{3}$O$_{7-\delta}$ superlattices. \textit{Phys. Rev. B} \textbf{67}, 214511 (2003).

\bibitem{Pena2004} Pe\~na, V. \textit{et al.} Coupling of superconductors through a half-metallic ferromagnet: Evidence for a long-range
proximity effect. \textit{Phys. Rev. B} \textbf{69}, 224502 (2004).

\bibitem{Hoppler2009} Hoppler, J. \textit{et al.} Giant superconductivity-induced modulation of the ferromagnetic magnetization in a cuprate-manganite superlattice. \textit{Nature Mater.} \textbf{8}, 315-319 (2009).

\bibitem{Kalcheim2011} Kalcheim, Y., Kirzhner, T., Koren, G. \& Millo, O. Long-range proximity effect in La$_{2/3}$Ca$_{1/3}$MnO$_{3}$/(100)YBa$_{2}$Cu$_{3}$O$_{7-\delta}$ ferromagnet/superconductor bilayers: Evidence for induced triplet superconductivity in the ferromagnet. \textit{Phys. Rev. B} \textbf{83}, 064510 (2011).

\bibitem{Visani2012} Visani, C. \textit{et al.}
Equal-spin Andreev reflection and long-range coherent transport in high-temperature superconductor/halfmetallic ferromagnet junctions.
\textit{Nature Phys.} \textbf{8}, 539-543 (2012).

\bibitem{Hoffmann2005} Hoffmann, A. \textit{et al.} Suppressed magnetization in La$_{0.7}$Ca$_{0.3}$MnO$_{3}$/YBa$_{2}$Cu$_{3}$O$_{7-\delta}$ superlattices. \textit{Phys. Rev. B} \textbf{72}, 140407(R) (2005).

\bibitem{Stahn2005} Stahn, J. \textit{et al.} Magnetic proximity effect in perovskite superconductor/ferromagnet multilayers. \textit{Phys. Rev. B} \textbf{71}, 140509(R) (2005).

\bibitem{Chakhalian2006} Chakhalian, J. \textit{et al.} Magnetism at the interface between ferromagnetic and superconducting oxides. \textit{Nature Phys.} \textbf{2}, 244-248 (2006).

\bibitem{Chakhalian2007} Chakhalian, J. \textit{et al.} Orbital Reconstruction and Covalent Bonding at an Oxide Interface. \textit{Science} \textbf{318}, 1114-1117 (2007).

\bibitem{Yunoki2007} Yunoki, S. \textit{et al.} Electron doping of cuprates via interfaces with manganites. \textit{Phys. Rev. B} \textbf{76}, 064532 (2007).

\bibitem{Chien2014} Chien, T. Y. \textit{et al.} Visualizing short-range charge transfer at the interfaces between ferromagnetic and superconducting oxides. \textit{Nature Commun.} \textbf{4}, 2336 (2013).

\bibitem{Satapathy2012} Satapathy, D. K. \textit{et al.} Magnetic Proximity Effect in YBa$_{2}$Cu$_{3}$O$_{7}$-La$_{2/3}$Ca$_{1/3}$MnO$_{3}$  and YBa$_{2}$Cu$_{3}$O$_{7}$-LaMnO$_{3+\delta}$ Superlattices. \textit{Phys. Rev. Lett.} \textbf{108}, 197201 (2012).

\bibitem{Heinze2012} Heinze, S. \textit{et al.}
Thermoelectric properties of YBa$_{2}$Cu$_{3}$O$_{6+\delta}$--La$_{2/3}$Ca$_{1/3}$MnO$_3$ superlattices. \textit{Appl. Phys. Lett.}  \textbf{101}, 131603 (2012).

\bibitem{Fink2013} Fink, J., Schierle, E. Weschke, E. \& Geck, J. Resonant elastic soft x-ray scattering. \textit{Rep. Progr. Phys.} \textbf{76}, 056502 (2013).

\bibitem{Liang2006} Liang, R. \textit{et al.} Evaluation of CuO$_2$ plane hole doping in YBa$_2$Cu$_3$O$_{6+x}$ single crystals. \textit{Phys. Rev. B} \textbf{73}, 180505 (2006).

\end{thebibliography}
\end{document}